\documentclass[usenatbib]{mn2e}
\usepackage{amsmath, epsfig, natbib}

\begin{document}

\title[Effect of gas on spiral pattern speed]
{Dynamical effect of gas on spiral pattern speed in galaxies}
\author[S. Ghosh  and C.J. Jog ]
       {Soumavo Ghosh$^{1}$\thanks{E-mail : soumavo@physics.iisc.ernet.in}, and
        Chanda J. Jog$^{1}$\thanks{E-mail : cjjog@physics.iisc.ernet.in}\\
$^1$   Department of Physics,
Indian Institute of Science, Bangalore 560012, India \\
}

\maketitle
\begin{abstract}
In the density wave theory of spiral structure, the grand-design two-armed spiral pattern is taken to rotate rigidly in a  galactic disc with a constant, definite pattern speed. The observational measurement of the pattern speed of the spiral arms, though difficult, has been achieved in a few galaxies such as NGC 6946, NGC 2997, and M 51 which we consider here. We examine whether the theoretical dispersion relation permits a real solution for wavenumber corresponding to a stable wave, for the observed rotation curve and the pattern speed values. We find that the disc when treated to consist of stars alone, as is usually done in literature, does not generally support a stable
density wave for the observed pattern speed. Instead the inclusion of the low velocity dispersion component, namely, gas, is essential to 
obtain a stable density wave. Further, we obtain a theoretical range of allowed pattern speeds that correspond to a stable 
 density wave at a certain radius, and check that for the three galaxies considered,  
the observed pattern speeds fall in the respective prescribed range. The inclusion of even a small amount ($\sim ~ 15 \%$) of gas by mass fraction in the galactic disc is shown to have a significant dynamical effect on the dispersion relation and hence on the pattern speed that is likely to be seen in a real, gas-rich spiral galaxy.
\end{abstract}


\section{Introduction}
Various surveys on galaxy morphology have revealed that the spiral structure is mainly of two types: flocculent structure,  and  grand-design spiral structure \citep{Elm11}. According to the density wave theory, the grand-design spiral pattern is a density wave that  rotates rigidly in a galactic disc, and is maintained by the disc gravity  \citep{LS64,LS66}, also see  \citet{DOBA14} for a recent review.

The determination of pattern speed is of particular interest in galactic dynamics as it sets the location of resonance points where angular momentum transport is believed to occur, thus it has direct implications for the secular evolution of the galactic disc \citep{LYKA72}. Over the years, several techniques have been devised to measure the pattern speed of these spiral structures. A common technique that is used is based on the assumed knowledge of the location of the resonance points \citep{BuCs96}, which  involves understanding the behaviour of stars and gas at these resonance points. Thus, estimation of the resonance points from the observed surface brightness profiles, will give the value of pattern speed \citep{PD97}. 
Another approach relies on information regarding the sign reversal of the radial streaming motion across the corotation. This can be done by studying a strip covering the kinematical minor-axis, where the radial component of the velocity has a non-zero projection along the line of sight. This method requires the position angle of the strip to be known accurately, which is not easy to check observationally because warps, often present in the outer parts of a galaxy, twist the location of the position angles.
A related method, first proposed by \citet{CAN93}, employs the change of sign of radial streaming motion across corotation and also takes account of the geometric phase values. This technique has been applied successfully first to NGC~4321 \citep{Sem95}. Also, the pattern speed can be estimated from the observed azimuthal age-gradient of the young stellar complexes which are  seen to be associated with the spiral arms \citep{GP98}. This technique has been applied to NGC~2997 to measure its pattern speed \citep{GD09}. Other popular method is the Tremaine-Weinberg method (hereafter, the TW method) which requires no specific dynamic model and predicts the value of pattern speed from kinematic measurements only \citep{TW84}. In the past, the TW method has been used to deduce the pattern speed of bars \citep{MK95,COR03,MA06} and spiral structures \citep{Fat07}. Each of the above methods involves different possible sources of errors or inaccuracies \citep[see e.g. the discussion in][]{Jun15}.

Spiral galaxies also contain a certain amount of interstellar gas whose fraction varies with their Hubble type \citep[e.g
][]{YoSc91,BM98}. The role of gas has been studied in various contexts in galactic dynamics, and it has been shown that the low velocity dispersion component, namely, gas has a significant effect on stability against both local axisymmetric \citep{JS84a,JS84b,BR88,Jog96,Raf01} and non-axisymmetric \citep{Jog92} perturbations.

The longevity of a density wave was questioned by \citet{Too69} who showed that a wavepacket of density waves would propagate radially with a group velocity $c_{\rm g}(R)$ = $\partial \omega(k, R) / \partial k$, where $\omega$ and $k$ are the frequency and the wavenumber, respectively, and $R$ is the radius; and the dependence of $\omega$ on $k$ is determined by the corresponding dispersion relation. This results in winding up of the wavepacket in a time-scale of about $10^9$ years. A recent work by \citet{GJ15} showed that the inclusion of gas in the disc makes the group transport slower by a factor of few, thus allowing the pattern to persist for a longer time-scale. They also showed that for the observed pattern speed of 18 km s$^{-1}$ kpc$^{-1}$ \citep{Sie12} and for assumed values of Toomre Q-parameters for our Galaxy, the disc when modelled as a stars-alone case, does not give a stable wave solution. Instead, one needs to invoke gas in order to get a stable density wave solution for the observed pattern speed.

In this paper, we study this in more detail and show this to be a general result: for this we study three external galaxies, NGC~6946, NGC~2997, and M~51 (NGC~5194)
for which the observational values of the pattern speeds for the spiral structure and the rotation curves are available in the literature. We first treat the galactic disc as comprised only of stars and from the dispersion relation we obtain the lowest possible value of the dimensionless frequency (see \S~2.2 for details) for which stars-alone will allow a stable wave. We next include gas on an equal footing with stars and follow the same procedure except for a two-component dispersion relation. We find that, at a radius equal to two disc scale-lengths and for an assumed set of Toomre Q parameter values, the stars-alone cases for NGC~6946 and NGC~2997 do not support a stable wave while for M~51, the stars-alone case marginally supports a stable density wave for the observed pattern speed. One has to include a gas fraction appropriate for the galaxies considered here, to get a stable density wave corresponding to the observed pattern speed.
As a check, we also varied the parameters considered, covering a reasonable range of values, and confirmed the validity of this finding. Also, based on these calculations, we derive a range of allowed pattern speeds that yields a stable density wave at a given radius of a galaxy.  We apply this method to these three galaxies, and find that the observed pattern speed values indeed fall in this prescribed range.

\S~2 describes the formulation of the problem while \S~3 presents the results. \S~4 and \S~5 contain the discussion and conclusion, respectively.

\section{Formulation of the problem}
We treat the galactic disc as a gravitationally coupled two-component (star plus gas) system, where the stars are treated as a collisionless system and characterized by the surface density $\Sigma_{\rm 0s}$ and the one-dimensional velocity dispersion $\sigma_{\rm s}$. The gas is treated as a fluid, characterized by the surface density $\Sigma_{\rm 0g}$ and a one-dimensional velocity dispersion or sound speed $\sigma_{\rm g}$.
Note that in any real galaxy, gas is seen to be present in both atomic (HI) and molecular (H$_2$) hydrogen form, having different surface density \citep{YoSc91,BB12} and velocity dispersion profiles \citep{Tam09}. For the sake of simplicity, here we treat gas as a single-component and study its effect on the dispersion relation.
 The galactic disc is taken to be infinitesimally thin and the pressure acts only in the disc plane. In other words, we are interested in gravitational perturbations in the disc plane only. We use the cylindrical co-ordinates (R, $\phi$, z).\\
\subsection{Dispersion relation in the WKB limit}
Consider the above system being perturbed by linear perturbations of the type  exp[${i (kr - \omega t)}$], where $k$ is the wavenumber and $\omega$ is the frequency of the perturbation.
For such a system, the dispersion relation in the WKB (Wentzel - Kramers - Brillouins) limit or the tightly wound case is \citep{GJ15}
\begin{equation}
\frac{2\pi G \Sigma_{\rm 0s} |k|F\Big(\frac{\omega-m \Omega}{\kappa},\frac{k^2\sigma^2_s}{\kappa^2}\Big)}{\kappa^2-(\omega-m\Omega)^2}+\frac{2\pi G \Sigma_{\rm 0g} |k|}{\kappa^2-(\omega-m\Omega)^2+\sigma^2_gk^2} = 1
\end{equation}
\noindent where $F$ is the reduction factor which physically takes into account the reduction in $\omega^2$ arising due to the velocity dispersion of stars. The functional form of $F$ is as given in \citet{BT87}. Here $\Omega$ is the angular frequency and $\kappa$ is the local epicyclic frequency at a given radius.

After some algebraic simplifications \citep [for details see][]{GJ15}, the dispersion relation (equation 1) reduces to
\begin{eqnarray}\nonumber
~~~~~~~~(\omega-m\Omega)^2=\frac{1}{2}\{(\alpha_s+\alpha_g)- [(\alpha_s+\alpha_g)^2-\\\nonumber
~~~~~~~~~4(\alpha_s\alpha_g-\beta_s\beta_g)]^{1/2}\}\\
\end{eqnarray}
where,
\begin{eqnarray}\nonumber
\alpha_s = \kappa^2-2\pi G\Sigma_{\rm 0s}|k| F\Big(\frac{\omega-m \Omega}{\kappa},\frac{k^2\sigma^2_s}{\kappa^2}\Big)\\\nonumber
\alpha_g = \kappa^2-2\pi G\Sigma_{\rm 0g}|k|+k^2{\sigma}^2_g\\\nonumber
\beta_s = 2\pi G\Sigma_{\rm 0s}|k|F\Big(\frac{\omega-m \Omega}{\kappa},\frac{k^2\sigma^2_s}{\kappa^2}\Big)\\\nonumber
\beta_g = 2\pi G\Sigma_{\rm 0g}|k| \\
\end{eqnarray}
Now we define two dimensionless quantities, $s$, the dimensionless frequency and $x$, the dimensionless wavenumber as\\
\begin{eqnarray}
s=({\omega-m\Omega})/{\kappa} =   {m(\Omega_p - \Omega})/{\kappa}, \: \:  x = k / k_{crit}
\end{eqnarray}
\noindent where $k_{crit} = \kappa^2 / 2 \pi G  (\Sigma_{\rm 0s} + \Sigma_{\rm 0g})$.
Substituting equation (4) in equation (2), we get the dimensionless form of the dispersion relation as \citep[see equations (7) \& (8) in][]{GJ15}:
\begin{eqnarray}
s^2=\frac{1}{2}[(\alpha'_s+\alpha'_g)-\{(\alpha'_s+\alpha'_g)^2-4(\alpha'_s\alpha'_g-\beta'_s\beta'_g)\}^{1/2}]
\label{stargas-disp}
\end{eqnarray}
where,
\begin{eqnarray}\nonumber
\alpha'_s=1-(1-\epsilon)|x|F(s,\chi) \\\nonumber 
\alpha'_g=1-\epsilon |x|+\frac{1}{4}Q^2_g\epsilon^2x^2 \\\nonumber
\beta'_s=(1-\epsilon)|x|F(s,\chi) \\\nonumber
\beta'_g = \epsilon |x|\\
\end{eqnarray}
and, $\chi$= ${k^2\sigma^2_s}/{\kappa^2}$ = $0.286 Q_s^2 (1-\epsilon)^2 x^2$.  
The three dimensionless parameters $Q_{\rm s}$, $Q_{\rm g}$, and $\epsilon$ are respectively the Toomre $Q$ factors for stars as a collisionless system  $Q_{\rm s}$(=$\kappa \sigma_s /(3.36 G \Sigma_{0s})$), and for gas $Q_{\rm g}$ = ($\kappa \sigma_g /(\pi G \Sigma_{0g})$) \citep{Too64}, 
and $\epsilon$ =${\Sigma_{0g}}/( {\Sigma_{0s}+\Sigma_{0g}})$ is the gas mass fraction in the disc.

Similarly, the one-component analog of this dispersion relation is \citep{BT87}:
\begin{equation}
s^2= 1-|x|F\Big(\frac{\omega-m \Omega}{\kappa},\frac{k^2\sigma^2_s}{\kappa^2}\Big)
\label{onefluid-disp}
\end{equation}
The dispersion relations (equation 7 for stars-alone and equation 5 for the two-component case) provide the information about how the dimensionless frequency $s$ varies locally with respect to the dimensionless wavenumber $x$.  Note that the equations (5) and (7) are symmetric with respect to both $s$ and $x$, hence we consider only their absolute values throughout this paper. It is clear that the absolute value of $s$ ($|s|$) ranges from 0 to 1 , where $s=0$ yields the position of corotation (hereafter CR) and $|s| =1$ gives the positions of Lindblad resonances \citep{BT87}.

 It has been shown that for any $Q_s> 1$, there exists a zone between Lindblad resonance and corotation, known as the forbidden region where the dispersion relations (equations 5 and 7) have no real solution for $|k|$. Hence the corresponding density wave solutions, that fall in that range, will be evanescent, i.e. they will have complex wavenumber $k$, and they will either decay or grow exponentially \citep{BT87}. A similar result holds when a two-component system has real $|s|$ solution.

A wavepacket, starting from the long-wave branch of the dispersion relation ($|x| < 1$), travels inward with a negative group velocity and gets reflected from the edge of the forbidden region near the CR, and then starts travelling radially outward in the short-wave branch ($|x|> 1$) with a positive group velocity (see Fig. 1) before finally being absorbed at a large wavenumber by a process similar to Landau damping \citep{BT87}.
\subsection {Method}
Note that, in the density wave theory, the pattern speed of the grand-design two-armed spiral structure, $\Omega_p$ (= $\omega$/m; m = 2 here for spirals) is treated as a free parameter \citep[e.g.,][]{LS66}, and its value is obtained from observations. Thus for a given observed value of $\Omega_ p$ and from the observed rotation curve, the dimensionless frequency  $|s|$ ($=m|\Omega_p- \Omega|/\kappa$) (equation 4), has a definite value at a certain radius R, say $|s|_{\rm obs}$. Therefore, for getting a stable density wave, the observationally found $|s|_{\rm obs}$ value should fall in the allowed range of $|s|$, derived from the dispersion relation, both obtained at the same radius. In other words, the horizontal line $|s|$ = $|s|_{\rm obs}$ should cut the plot of $|s|$ versus $|x|$ at that radius, to yield a stable wave solution (i.e. a real solution for $|k|$), otherwise it will give an evanescent density wave solution. 

\begin{figure}
\centering
\includegraphics[height=2.5in,width=3.5in]{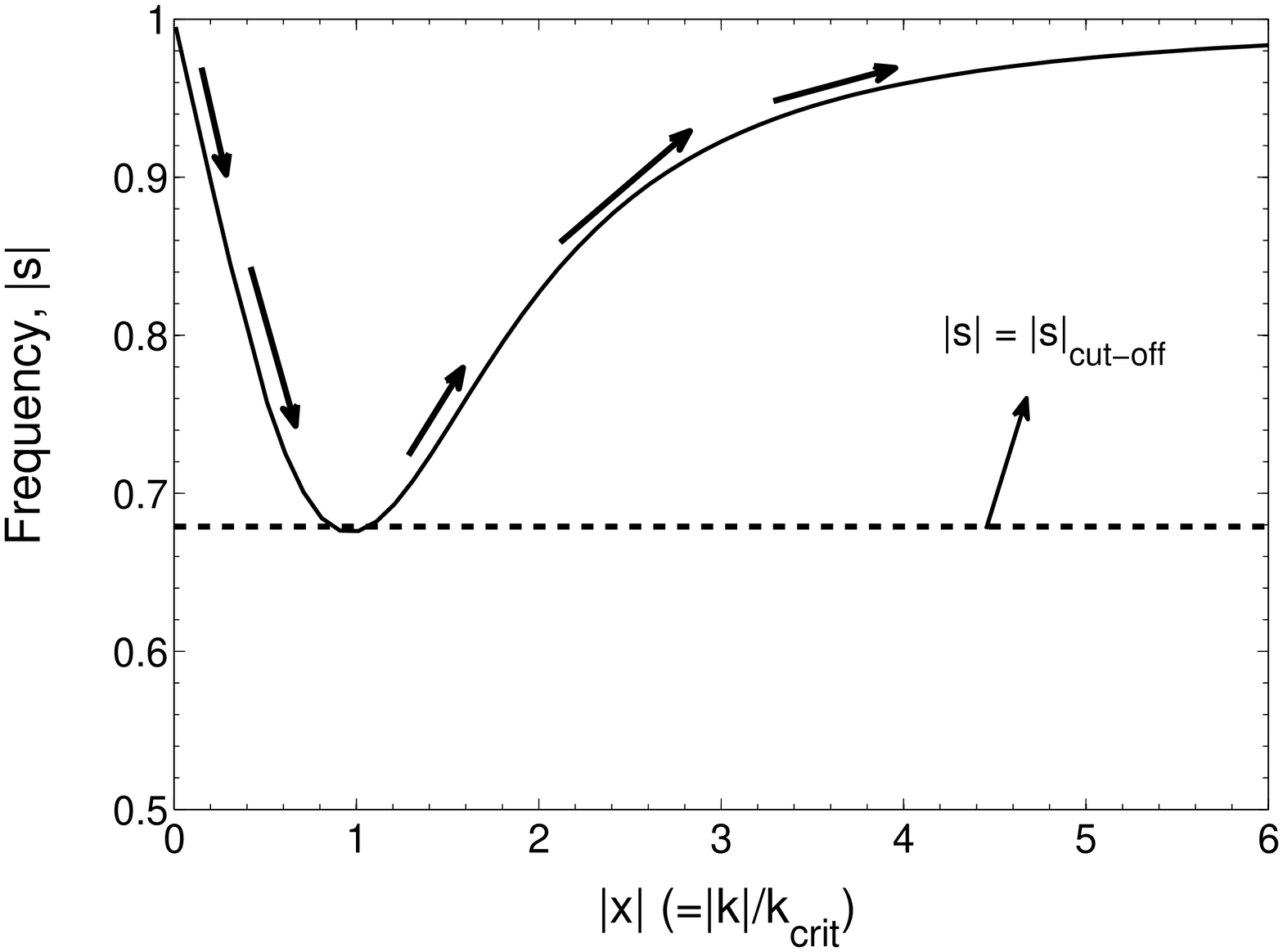}
\caption{The dimensionless frequency $|s|$ vs. the dimensionless wavenumber $|x|$ from the dispersion relation for the stars-alone case (equation 7), plotted for $Q_s =1.7$. Since the dispersion relation is symmetric in both $s$ and $x$, only their absolute values are shown here. The arrows denote the direction of propagation of a typical density wavepacket. The horizontal line $|s|$ = $|s|_{\rm cut-off}$ indicates the lowest possible value of $|s|$, for which one can get a stable wave solution.}
\label{fig1}
\end{figure}

We define $|s|_{\rm cut-off}$ as the lowest possible value of $|s|$ for which one is able to get a stable wave solution from the dispersion relation  at a given radius $R$. We set $|s|_{\rm cut-off}$ as the lowest $|s|$-value, where the plot of the dispersion relation turns around (equation (5) or equation (7), whichever is applicable). In other words, $|s|_{\rm cut-off}$ indicates the edge of the forbidden region. A typical example of how $|s|_{\rm cut-off}$ is obtained from a dispersion relation for a stars-alone case, is shown in Fig.~{\ref{fig1}}. For the sake of illustration, we assume $Q_{\rm s}$ = 1.7, as observed in the solar neighbourhood \citep{BT87}.

Thus, if the following inequality holds for a certain radius $R$: 
\begin{equation}
|s|_{obs} \ge |s|_{\rm cut-off}
\end{equation}
then one can say that the observed pattern speed will give a stable density wave, and the dispersion relation will have real solution for $|k|$.
Throughout this paper, $|s|_{\rm cut-off}$ will be used for making quantitative statements
regarding the existence of stable solutions.
\section{Results}
\subsection{Input parameters}
We consider three galaxies for which the observational values for the input parameters, namely the pattern speed and the rotation curve, are available in the literature.
\subsubsection{NGC~6946}
This is a barred grand-design spiral galaxy with the Hubble type Scd. It has an angle of inclination $38^0$ \citep{CAR90,Boo07} and it is located at a distance of 5.5 Mpc \citep{KEN03}. The pattern speed of the main $m = 2$ gravitational perturbation i. e., the large oval and the two prominent spiral arms is measured to be $22_{-1}^{+4}$ km s$^{-1}$ kpc$^{-1}$ by \citet{Fat07}, using the TW method. They also derived the locations of different resonance points (see table 2 there). The HI rotation curve is taken from \citet{Boo07}. The exponential disc scale-length is measured as $1.9'$ or 3.3 kpc, where $1' = 1.75$ kpc \citep{CAR90}.

\subsubsection{NGC~2997}
This is a grand-design spiral galaxy of the Hubble type Sc \citep{MM81}. The pattern speed of the grand-design spiral structure is measured
to be  16 km s$^{-1}$ kpc$^{-1}$ by \citet{GD09}, by using the measurement of azimuthal age-gradient of newly formed stars. 
The rotation curve is taken from \citet{PET78}. The exponential disc scale-length is measured as 4.0 kpc \citep{GBP99}

\subsubsection{M~51 (NGC~5194)}
This is a face-on spiral galaxy of the Hubble type Sc and is located at a distance of 9.6 Mpc \citep{SanTamm75}, in close interaction with NGC~5195. The pattern speed of the spiral structure is measured to be 38 km s$^{-1}$ kpc$^{-1}$ by \citet{Zimm04}, using CO as a tracer and by applying the TW method. The rotation curve for M~51 is found to be steeply rising up to $\sim$ $R = 25''$ and then for R $>$ $25''$ it saturates to 210 km s$^{-1}$ \citep{RAND93}, where $1''=46.5$ pc, assuming a distance of 9.6 Mpc \citep{SanTamm75}. The observed rotation curve and the observed pattern speed together place the CR at a distance of 5.5 kpc. The exponential disc scale-length is measured in various wavelengths and is found to vary from 4.36 kpc in B-band to 3.77 kpc in R-band \citep{Beck96}. We took a mean value of 4.0 kpc for the present purpose.
\subsection{Stars-alone case}
We first investigate whether a disc, consisting only of stars, can support a stable density wave for the observed values of pattern speed in different galaxies. To do this, first we obtained $|s|_{\rm obs}$ 
at a radius of 2$R_{\rm d}$, $R_{\rm d}$ being the exponential disc scale-length, in each of these three galaxies.
The choice of 2R$_{\rm d}$ is made because the spiral structure is typically seen in the middle part of an optical disc whose size is $\sim$ 4-5 R$_{\rm d}$ \citep[e.g.,][]{BM98}. We found that the $|s|_{\rm obs}$ values for NGC~6946, NGC~2997, and M~51 are 0.38, 0.44, and 0.63 respectively.

Now to check whether the inequality given in equation (8) is satisfied or not, we need to calculate the $|s|_{\rm cut-off}$ value from the dispersion relation  (equation 7) for these galaxies. Note that the calculation of $|s|_{\rm cut-off}$ from the dispersion relation requires the knowledge of $Q_{s}$, and the values of $Q_{\rm s}$ at different radii for  a galaxy are not known observationally for most galaxies. We assume $Q_{\rm s}$ to be constant with $R$ for simplicity, but in reality its values will vary \citep[e.g., see][]{Jog14,GJ14}. Further, for a theoretical study, we varied the value of this radially-constant $Q_{\rm s}$ from $1.3$ to $2.0$, while $1.7$ is the typical value in the solar neighbourhood \citep{BT87}. The plot of the resulting $|s|_{\rm cut-off}$ versus $Q_{\rm s}$ is shown in Fig. 2. 
\begin{figure}
\centering
\includegraphics[height=2.5in,width=3.5in]{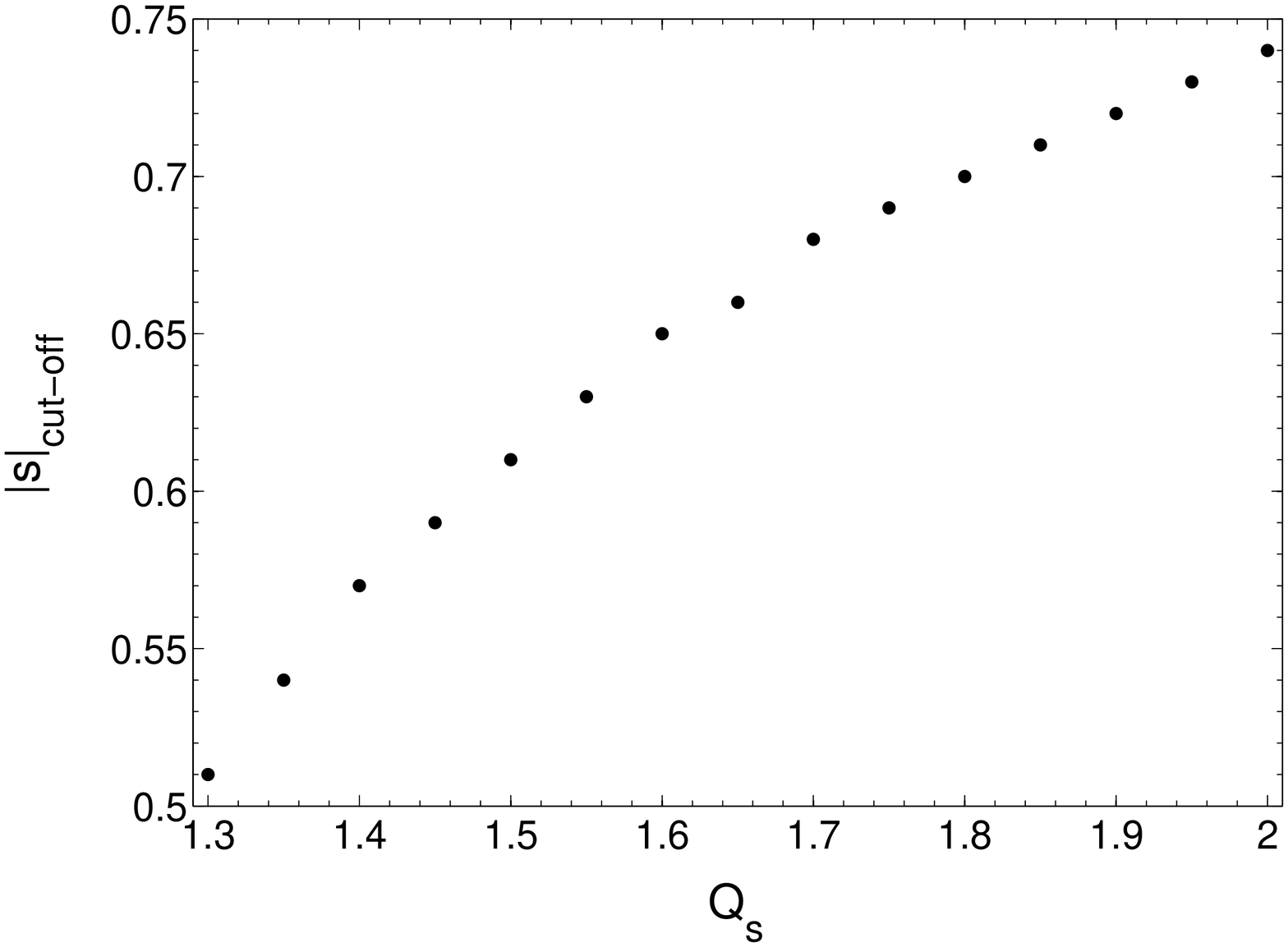}
\caption{$|s|_{\rm cut-off}$, the lowest values of the dimensionless frequency for which a stable density wave solution is possible, 
plotted as a function of $Q_s$, obtained from the one-component dispersion relation (equation 7). The resulting $|s|_{\rm cut-off}$ value shows a steady increase with $Q_{\rm s}$, thereby implying an increase in the forbidden region.}
\label{fig2}
\end{figure}
From Fig.~{\ref{fig2}}, we see that the $|s|_{\rm cut-off}$ value increases steadily with the increase of $Q_{\rm s}$, thereby implying a steady increase in the forbidden region around the CR as the $Q_{\rm s}$ value increases. 

On comparing the $|s|_{\rm obs}$ values that we obtained for both NGC~6946 and NGC~2997, with Fig.~{\ref{fig2}}, we find that the inequality given by equation (8) is not satisfied for any value of $Q_{s}$, considered here. Thus, the dispersion relation for a purely stellar disc does not yield a real solution for $|k|$ and hence it does not yield a stable density wave corresponding to the observed pattern speeds of both these galaxies. For M~51, the stars-alone case barely supports a stable density wave when $Q_{\rm s} =$ 1.5, but if it is set to a value larger than 1.5, then the stars-alone case no longer supports a stable density wave for the observed pattern speed.

\subsection{Stars plus gas case}
We next include the interstellar gas, which has a low velocity dispersion as compared to the stars, in the system, and study whether the inclusion of gas helps in getting a stable wave solution for the observed pattern speeds.
First we investigated how the values of $|s|_{\rm cut-off}$ change with different values for the three parameters $Q_{\rm s}$, $Q_{\rm g}$, and $\epsilon$. We varied $Q_{\rm s}$ from 1.3 to 2.0 and $\epsilon$ from 0.1 to 0.25. In each case we fix $Q_{\rm g}$, and then compute $|s|_{\rm cut-off}$ from equation (5), as a function of $Q_{\rm s}$, and repeat this procedure for different values of $\epsilon$. For comparison, we replotted the $|s|_{\rm cut-off}$ values for the stars-alone case, as a function of $Q_{\rm s}$. The result for $Q_{g} = 1.5$ is shown in Fig. 3. From Fig 3, it is clear that with the inclusion of more gas (i.e., higher $\epsilon$ value), the $|s|_{\rm cut-off}$ value steadily decreases. This holds true for the above mentioned range of $Q_{\rm s}$ values, at a given $Q_{\rm g}$. In other words, a larger value of gas fraction ($\epsilon$) helps to decrease the forbidden region around the CR, and thus allowing a higher range of permitted pattern speeds (for details see \S~3.4). For the other values of $Q_{\rm g}$ in the range of 1.4 to 1.8, we got a similar trend, hence we do not produce them here.
\begin{figure}
\centering
\includegraphics[height=2.5in,width=3.5in]{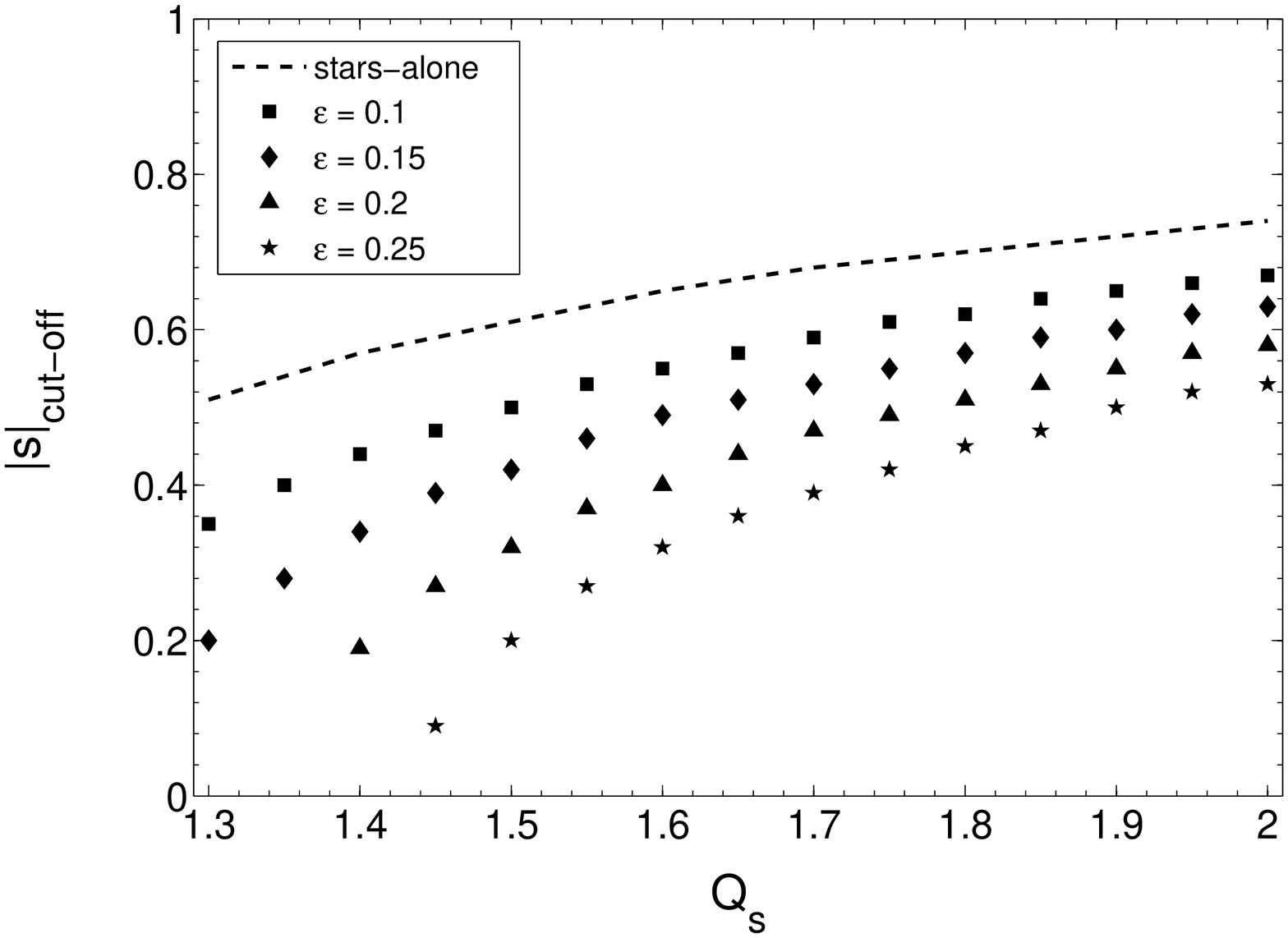}
\caption{$|s|_{\rm cut-off}$, the lowest values of the dimensionless frequency for which a stable density wave solution is possible for the two-component case, 
plotted as a function of $Q_s$, for different gas fractions ($\epsilon$) and $Q_{\rm g}$ = 1.5. The labels used for different gas fractions are as indicated in the legend. With increasing gas fraction, the $|s|_{\rm cut-off}$ decreases steadily for the whole range of $Q_{\rm s}$ considered here. Thus the extent of the forbidden region decreases with increasing gas fraction.}
\label{fig3}
\end{figure}

Now we investigate the dynamical effect of gas on the grand-design spiral structure for the three specific galaxies, chosen for this work. The value of $|s|_{\rm cut-off}$ is obtained from the dispersion relation for a two-component system (equation 5) for a set of values for the three
dimensionless input parameters ($Q_{\rm s}$, $Q_{\rm g}$, $\epsilon$). The values used are $Q_{\rm s} = 1.5$ and $Q_{\rm g} = 1.5$ for the stars plus gas case, and
$Q_{\rm s} = 1.5$ for the stars-alone case, with $\epsilon=0.25$ for NGC 6946 and $\epsilon=0.15$ for NGC 2997 and M~51, as typical for their Hubble types \citep[see fig. 5,][]{YoSc91}. Then we checked whether or not the value of $|s|_{\rm cut-off}$ obtained theoretically from the dispersion relation for the above input parameters, and the value of $|s|_{\rm obs}$ obtained from observations satisfy the inequality in equation (8). 
The results for NGC~6946, NGC~2997, and M~51 are shown in Fig.~4, Fig.~{\ref{fig5}}, and in Fig.~{\ref{fig6}}, respectively.

Since, the above values of $Q_{\rm s}$ and $Q_{\rm g}$ are chosen in a somewhat ad-hoc way, hence for each galaxy, we next study the variation in the dispersion relation for a reasonable range of $Q_{\rm s}$ and $Q_{\rm g}$ values, $Q_{\rm s}$ = 1.5, 1.6, 1.7 and $Q_{\rm g}$ = 1.4, 1.5, 1.6 and 1.7. The typical gas fraction value, chosen as per the Hubble type of any individual galaxy \citep[for details see][]{YoSc91} is kept constant. We found that, for a fixed $\epsilon$, there is a strong variation in the behaviour of the dispersion relation for different $Q_{\rm s}$ values, as compared to the different $Q_{\rm g}$ values.
A higher gas fraction ($\epsilon$) would be expected to change the results substantially, as suggested by \citet[see fig. 3 there]{JS84a}, but here we have kept $\epsilon$ constant as typical for a given Hubble type.
\begin{figure*}
    \centering
    \begin{minipage}{.32\textwidth}
        \centering
        \includegraphics[height=2.1in,width=2.3in]{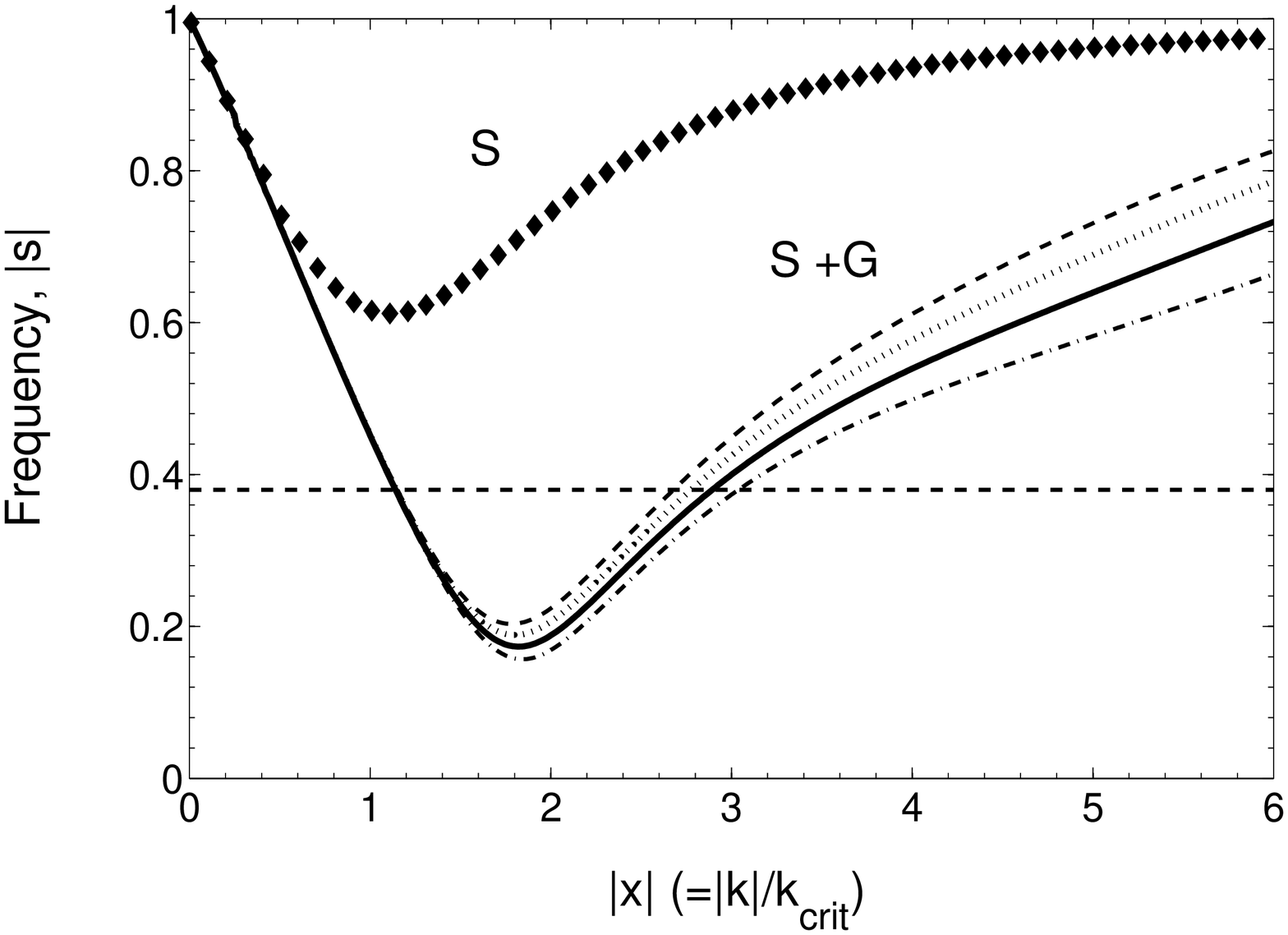}
       \vspace{0.2 cm}
	{\bf{(a)}}\\
    \end{minipage}
    \begin{minipage}{.32\textwidth}
        \centering
        \includegraphics[height=2.1in,width=2.3in]{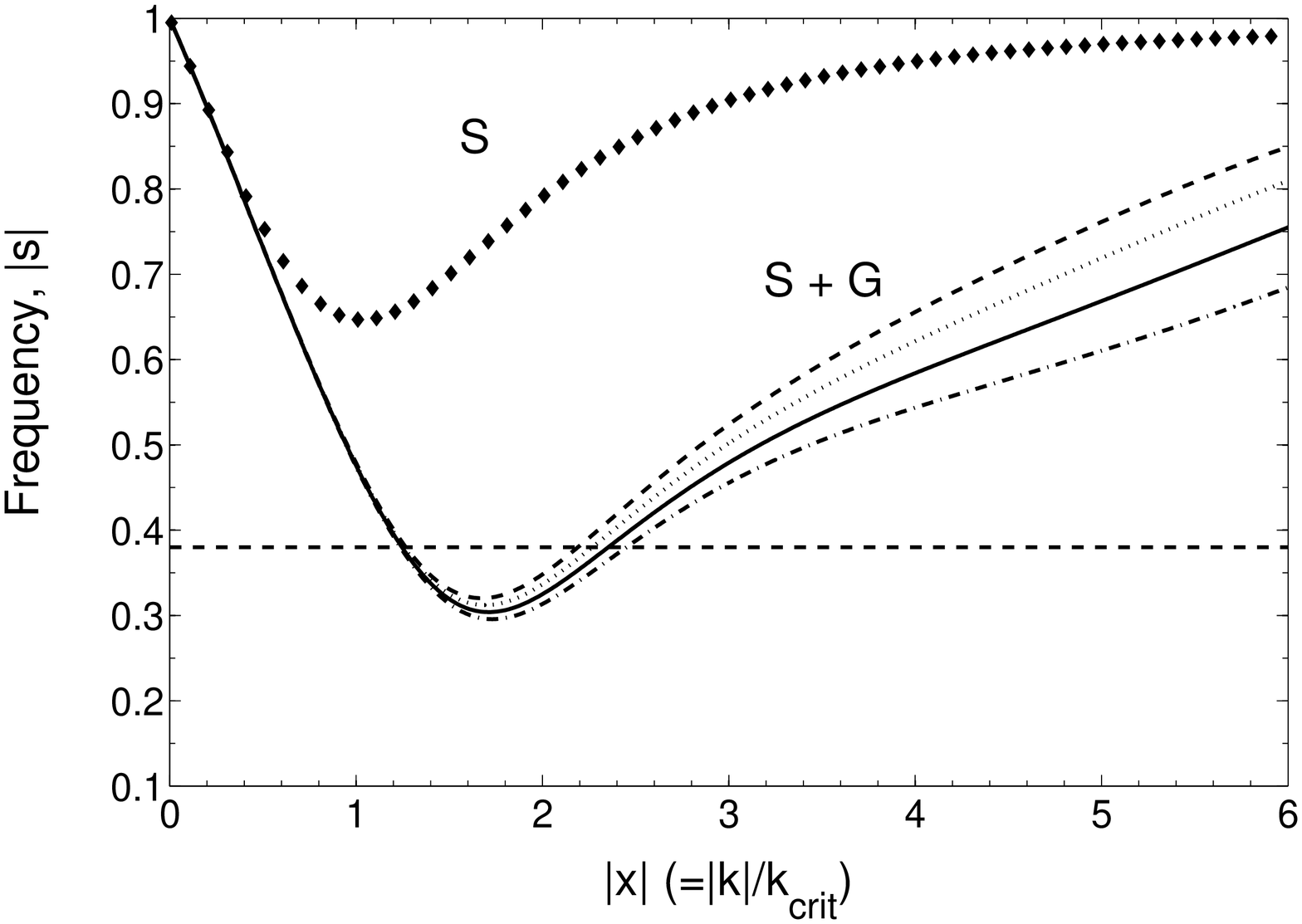}
        \vspace{0.2 cm}
	{\bf{(b)}}\\
    \end{minipage}
\begin{minipage}{.32\textwidth}
        \centering
        \includegraphics[height=2.1in,width=2.3in]{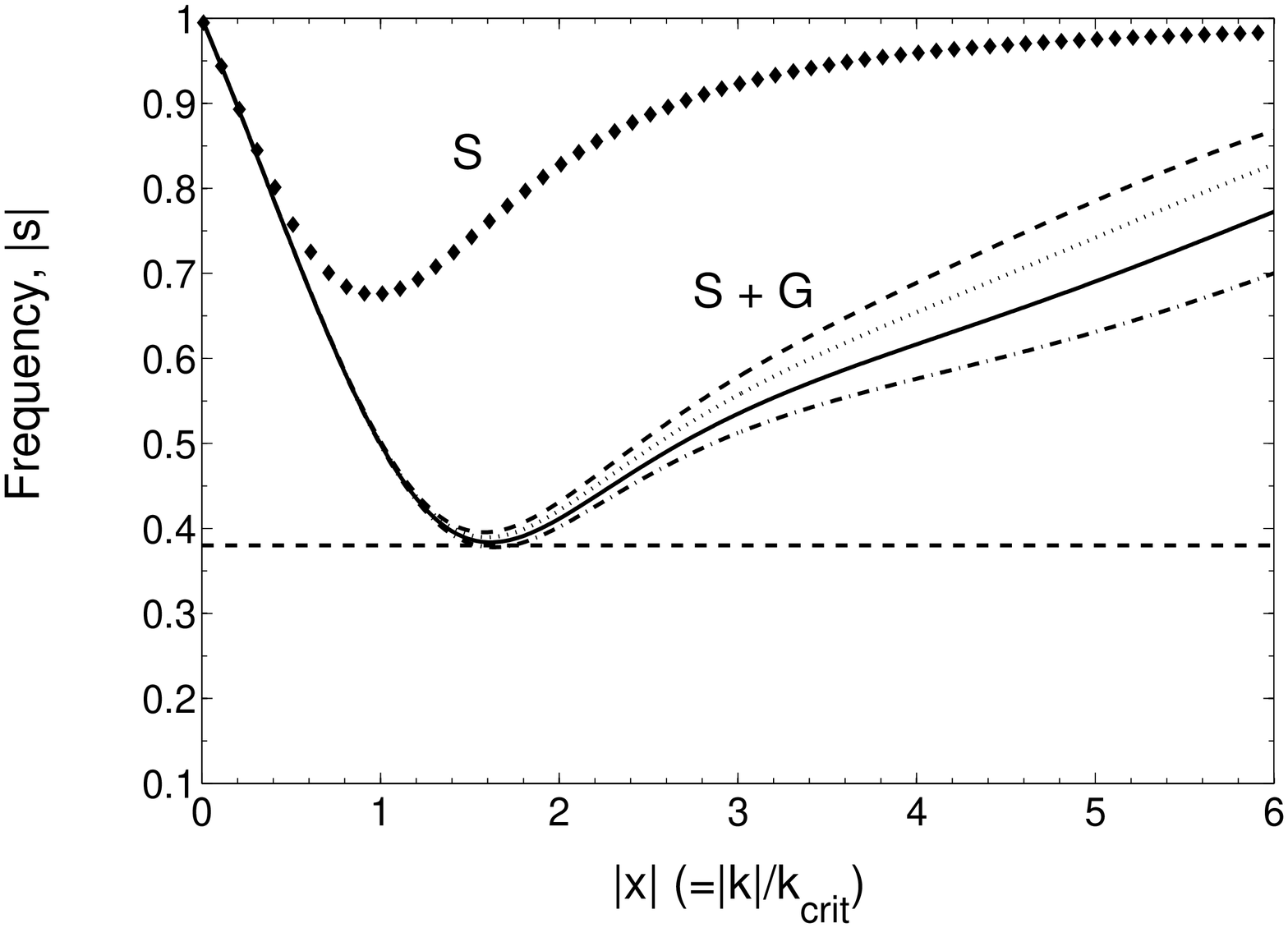}
        \vspace{0.2 cm}
	{\bf{(c)}}\\
    \end{minipage}
    \caption{{\it{NGC~6946}} : Dispersion relations for stars-alone (S) and stars plus gas (S + G) cases, plotted in a dimensionless form, for a range of $Q_{\rm s}$ and $Q_{\rm g}$ values, at R = 2R$_{\rm d}$. Panel (a) for $Q_{\rm s}=1.5$, panel (b) for $Q_{\rm s}=1.6$, and panel (c) for $Q_{\rm s}=1.7$. In each panel, $Q_{\rm g}$ is taken to be 1.4, 1.5, 1.6, and 1.7, successively. The corresponding dispersion relations are shown from bottom to top. The horizontal line indicates the value $|s|_{\rm obs}$, derived from the observed pattern speed and the rotation curve. Here in all the cases, the two-component case allows a real solution for $|k|$ and thus a stable density wave solution for the observed value of the pattern speed, but this is not true for the stars-alone case.}
    \label{fig4}
\end{figure*}

\begin{figure*}
    \centering
    \begin{minipage}{.32\textwidth}
        \centering
        \includegraphics[height=2.1in,width=2.3in]{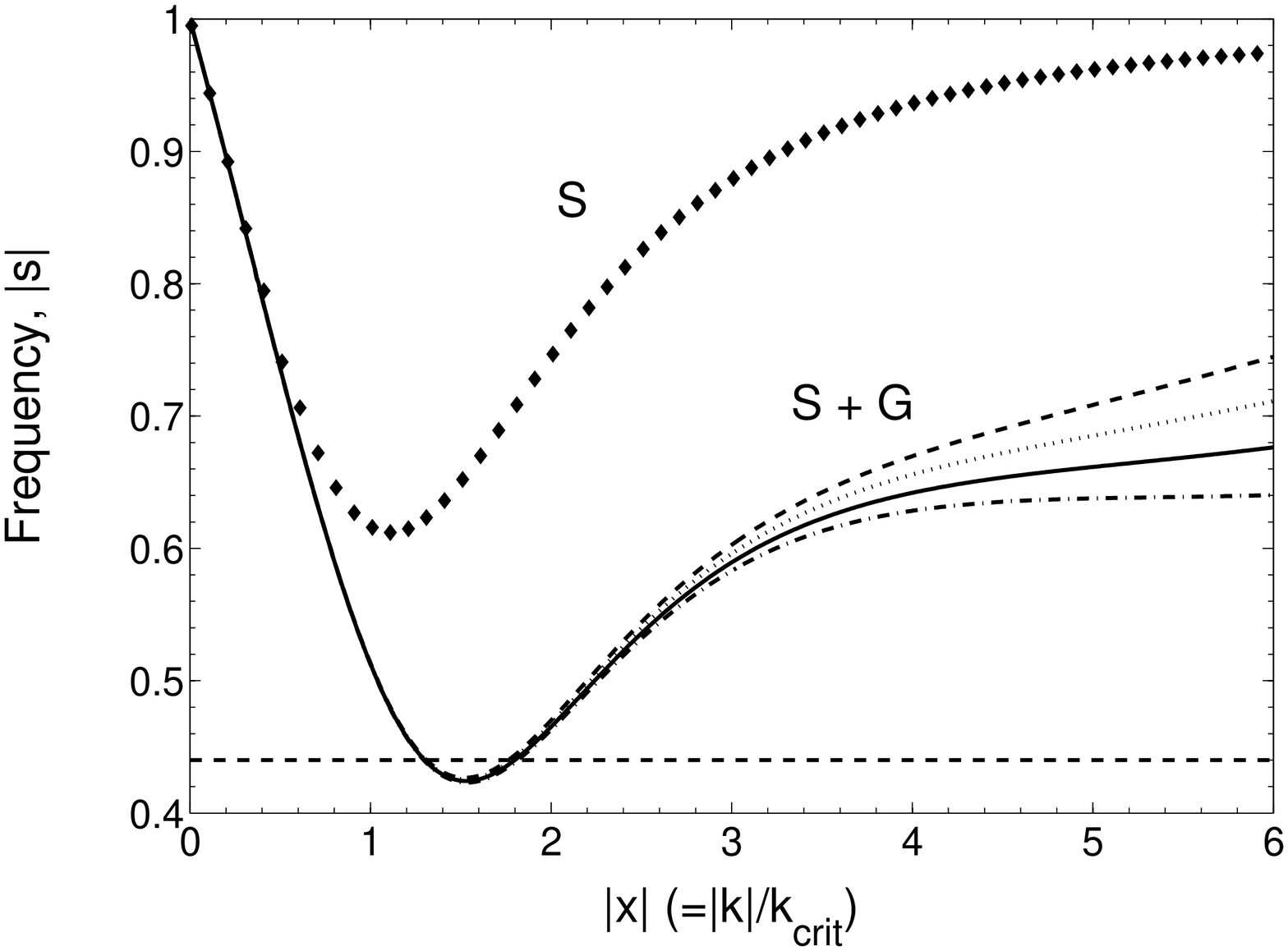}
       \vspace{0.2 cm}
	{\bf{(a)}}\\
    \end{minipage}
    \begin{minipage}{.32\textwidth}
        \centering
        \includegraphics[height=2.1in,width=2.3in]{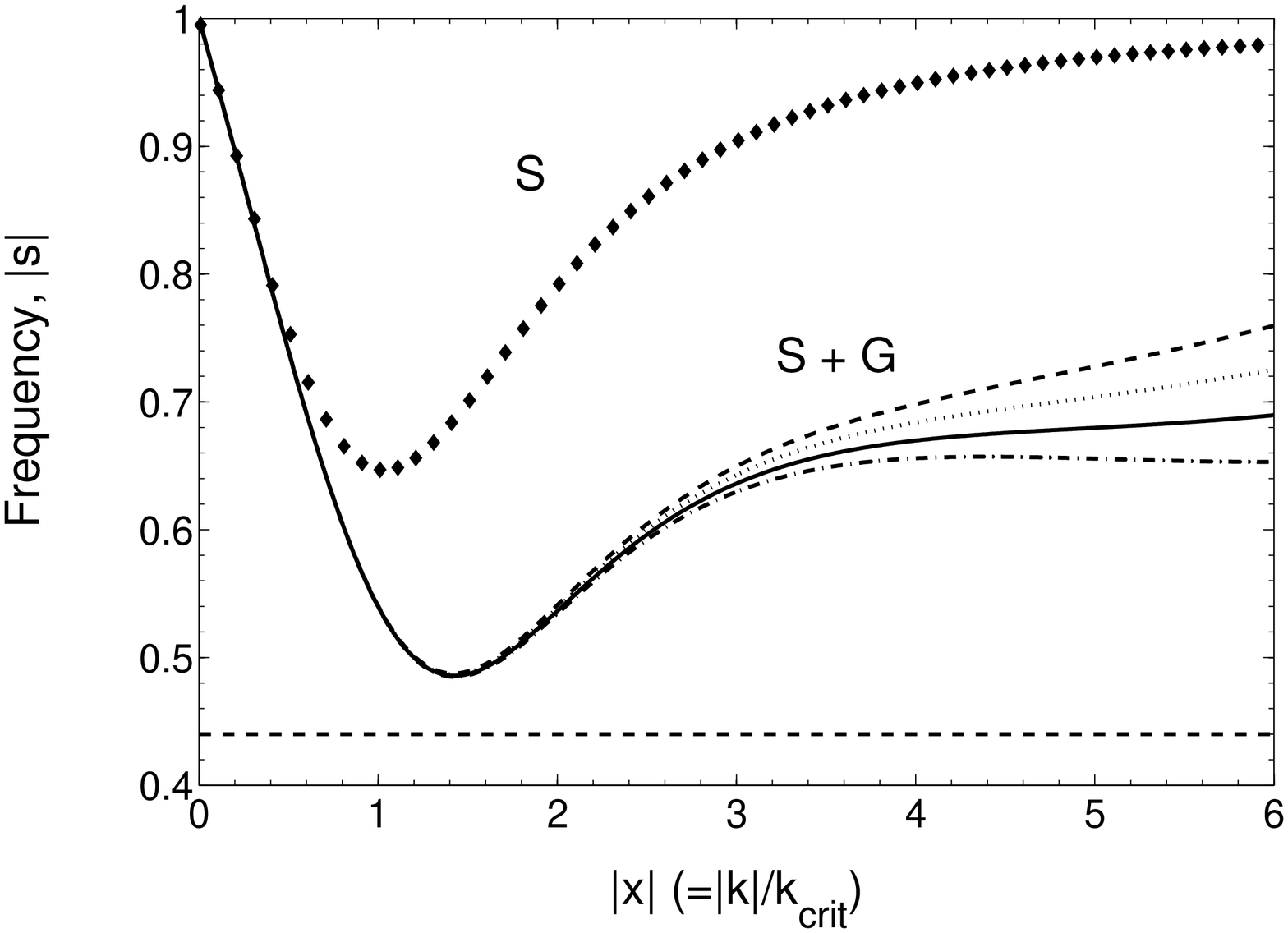}
        \vspace{0.2 cm}
	{\bf{(b)}}\\
    \end{minipage}
\begin{minipage}{.32\textwidth}
        \centering
        \includegraphics[height=2.1in,width=2.3in]{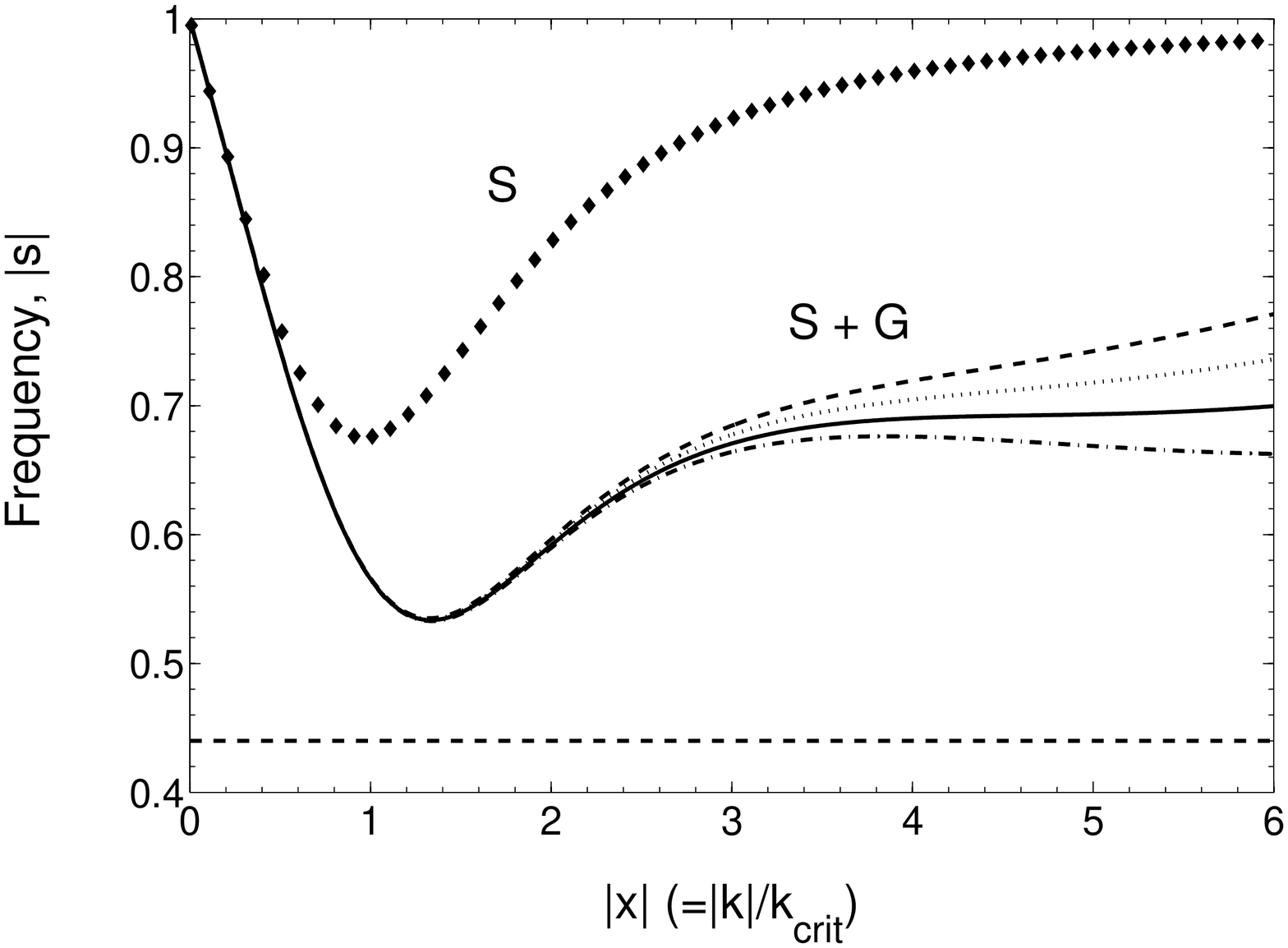}
        \vspace{0.2 cm}
	{\bf{(c)}}\\
    \end{minipage}
    \caption{{\it{NGC~2997}} : Dispersion relations for stars-alone (S) and stars plus gas (S + G) cases, plotted in a dimensionless form, for a range of $Q_{\rm s}$ and $Q_{\rm g}$ values, at R = 2R$_{\rm d}$. Panel (a) for $Q_{\rm s}=1.5$, panel (b) for $Q_{\rm s}=1.6$, and panel (c) for $Q_{\rm s}=1.7$. In each panel, $Q_{\rm g}$ is taken to be 1.4, 1.5, 1.6, and 1.7, successively. The corresponding dispersion relations are shown from bottom to top. The horizontal line indicates the value $|s|_{\rm obs}$, derived from the observed pattern speed and the rotation curve. Here for $Q_{\rm s} = 1.5$, the two-component case, but not the stars-alone case, allows a stable density wave solution while for other values of $Q_{\rm s}$, none of the stars-alone and two-component case allows a stable density wave for the observed value of the pattern speed.}
    \label{fig5}
\end{figure*}
\begin{figure*}
    \centering
    \begin{minipage}{.32\textwidth}
        \centering
        \includegraphics[height=2.1in,width=2.3in]{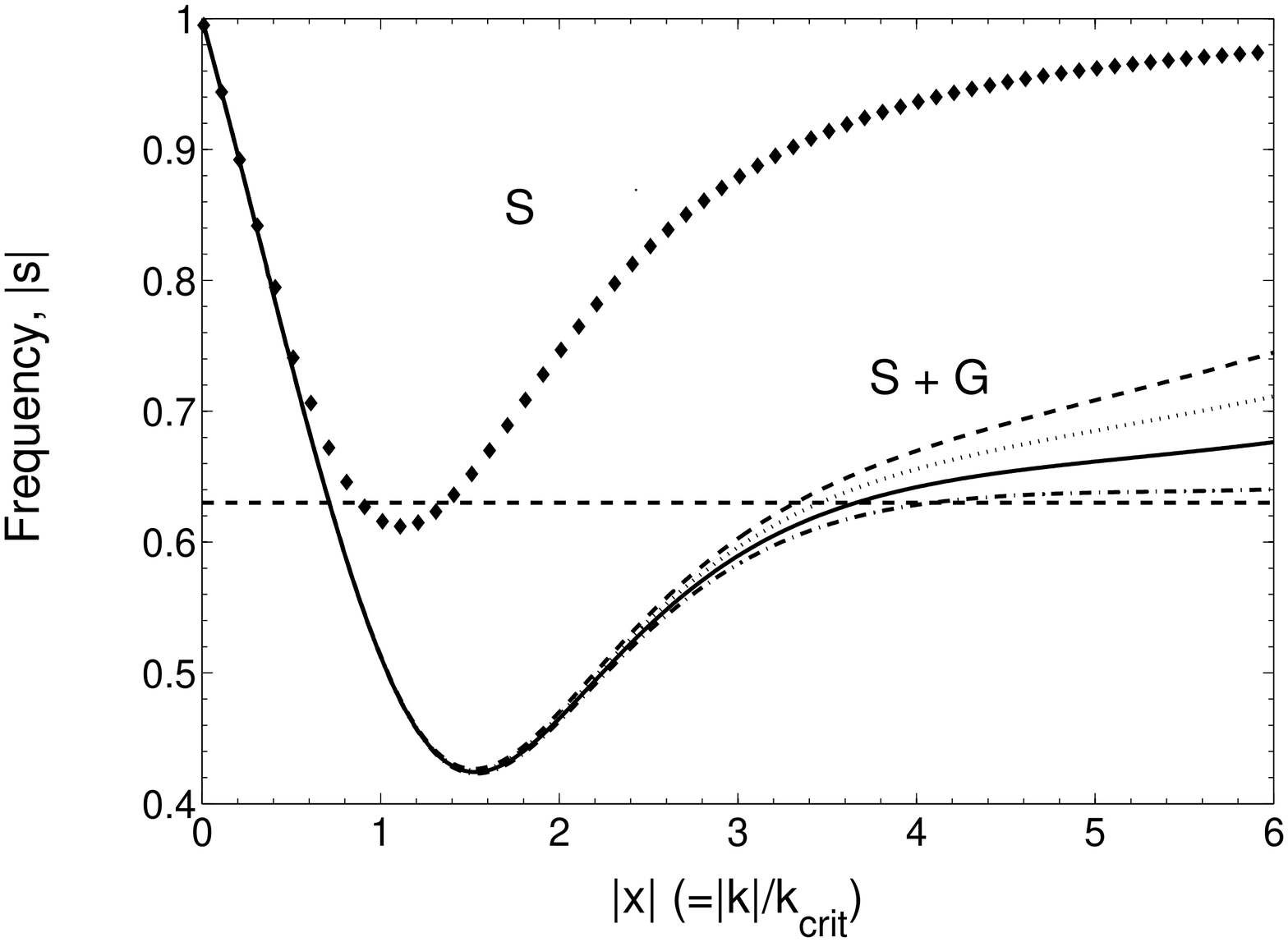}
       \vspace{0.2 cm}
	{\bf{(a)}}\\
    \end{minipage}
    \begin{minipage}{.32\textwidth}
        \centering
        \includegraphics[height=2.1in,width=2.3in]{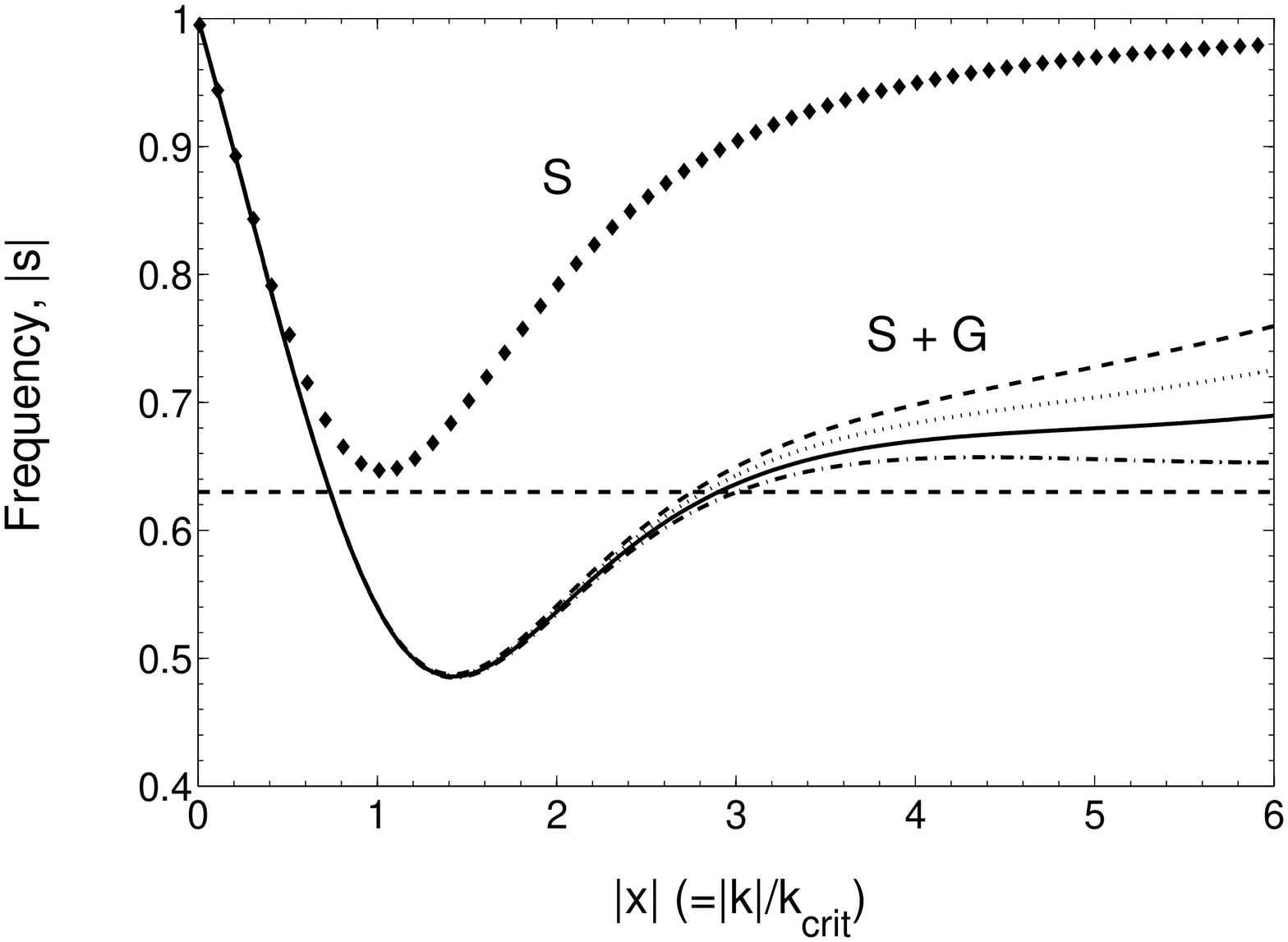}
        \vspace{0.2 cm}
	{\bf{(b)}}\\
    \end{minipage}
\begin{minipage}{.32\textwidth}
        \centering
        \includegraphics[height=2.1in,width=2.3in]{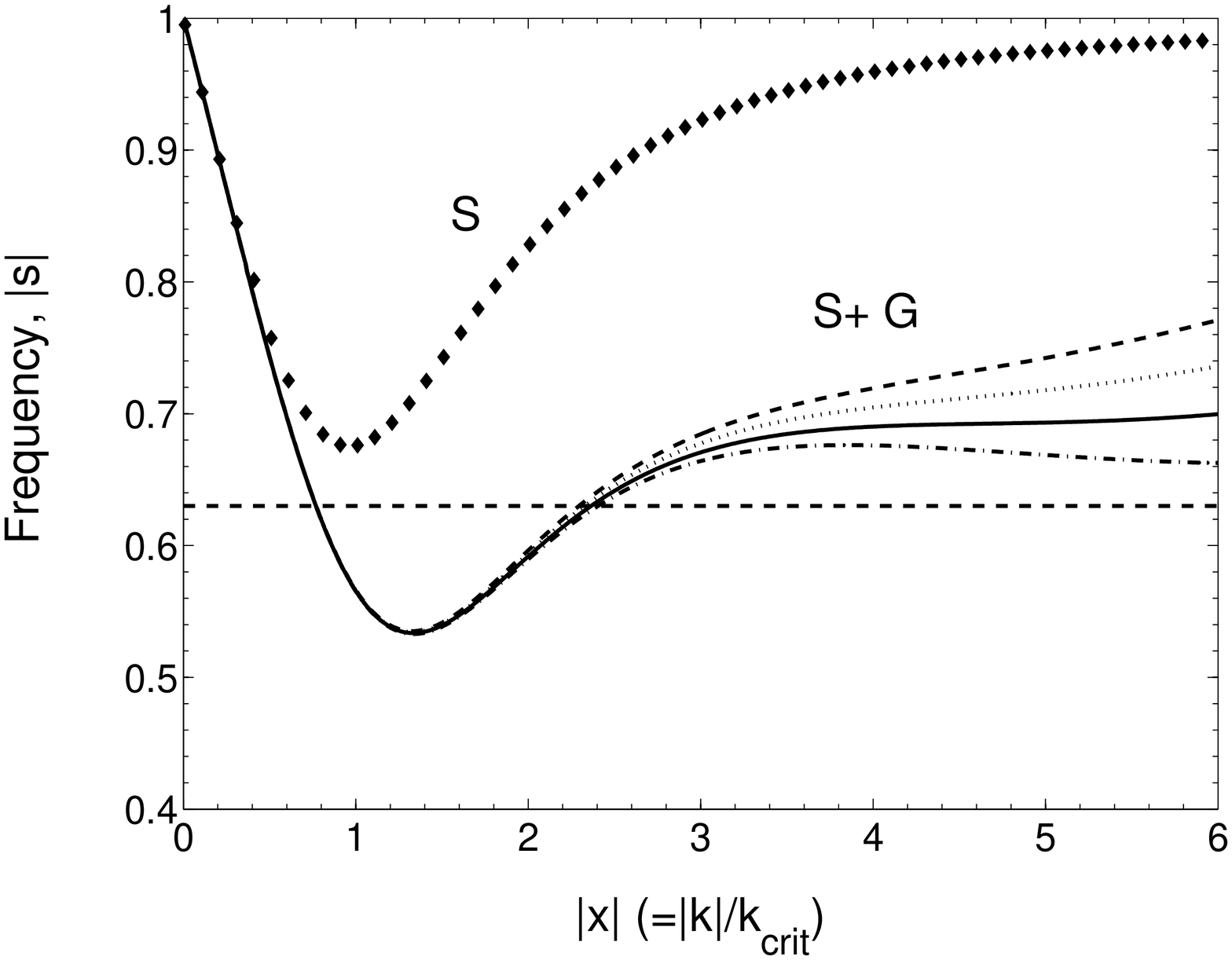}
        \vspace{0.2 cm}
	{\bf{(c)}}\\
    \end{minipage}
    \caption{{\it{M~51}} : Dispersion relations for stars-alone (S) and stars plus gas (S + G) cases, plotted in a dimensionless form, for a range of $Q_{\rm s}$ and $Q_{\rm g}$ values, at R = 2R$_{\rm d}$. Panel (a) for $Q_{\rm s}=1.5$, panel (b) for $Q_{\rm s}=1.6$, and panel (c) for $Q_{\rm s}=1.7$. In each panel, $Q_{\rm g}$ is taken to be 1.4, 1.5, 1.6, and 1.7, successively. The corresponding dispersion relations are shown from bottom to top. The horizontal line indicates the value $|s|_{\rm obs}$, derived from the observed pattern speed and the rotation curve. Here for $Q_{\rm s}=1.5$, both the two-component case and the stars-alone case, allow a stable density wave solution, but for larger values of $Q_{\rm s}$, only the two-component case allows a stable density wave for the observed value of pattern speed.}
    \label{fig6}
\end{figure*}
Fig.~{\ref{fig4}}, and Fig.~{\ref{fig6}}  show that for NGC~6946 and M~51, the $|s|_{\rm obs}$ value lies above the $|s|_{\rm cut-off}$ value for the star-gas system, thus the inequality given by equation (8) is satisfied, and this is true for the whole range of parameter space considered here. Thus the stars plus gas case allows a stable solution. For NGC~2997, the inequality given by equation (8) is satisfied only for $Q_{\rm s} = 1.5$. For a $Q_{\rm s}$ value higher than 1.5, even the addition of gas in the two-component system no longer admits a stable density wave (see Fig.~5). In contrast, for $Q_{\rm s} < 1.5$, the two component system admits a stable density wave solution. This can be seen from the result that as $Q_{\rm s}$ decreases, the $|s|_{\rm cut-off}$ for the stars plus gas also decreases, so that the forbidden region is smaller (see Fig.~{\ref{fig3}}).

 Hence, the galactic disc when treated as a gravitationally coupled stars plus gas system,
with the observed gas fraction,  allows a stable wave solution for the observed pattern speed, for most of the parameter range considered here. This is the main finding from this paper.

Interestingly, we see that for both NGC~6946 and NGC~2997 (Fig. 4 \& Fig.~5 (a)), the observed value of $|s|_{\rm obs}$ is close to the $|s|_{\rm cut-off}$ value which is obtained theoretically from the dispersion relation for the stars plus gas case (equation {\ref{stargas-disp}), but not the  $|s|_{\rm cut-off}$ value obtained for the stars-alone case (equation 7). Note that the curve for the dispersion relation for stars-alone case lies above that for the stars plus gas case, so if the pattern speed were such that the corresponding observed $|s|_{\rm obs}$ were to be  greater $|s|_{\rm cut-off}$ value for the stars-lone case it would also be greater than the cut-off value for the two-component case.
 Thus if a pattern speed value gives a stable density wave solution for a one-component case, it will also give so for the two-component case. Also the observed value of $|s|_{\rm obs}$ lies close to  the two-component $|s|_{\rm cut-off}$ value (but see M~51), and
this means that a galaxy seems to ``prefer'' to have a pattern speed that is indicated by the stars-plus-gas case for the observed gas fraction.

This can be explained as follows.
A joint two-component system is more unstable to the growth of perturbations or is closer to being unstable 
than the stars-alone case (Jog \& Solomon 1984), and this results in a lower cut-off $|s|_{\rm cut-off}$
for the two component case than the stars-alone case. If the galactic disc were subjected to perturbations having a range of values for pattern speeds (say as arising
due to a tidal interaction), then the perturbation most likely to be amplified in a galaxy is the one for which the dimensionless frequency $|s|_{\rm cut-off}$
has the lowest value, as this would correspond to the fastest growing perturbation.

This indicates that the inclusion of even 15 to 25 per cent gas fraction by mass has a non-trivial effect on the determination of the pattern speed that a galaxy is likely to have. 
The value of the pattern speed, along with the rotation curve, sets the locations of the Lindblad resonances in the disc. These are important in determining the secular evolution of galactic disc via angular momentum transport \citep{LYKA72}. 
Thus our work shows that the gas 
is important in setting the observed pattern speed in a galaxy and hence it is likely to play a crucial role in future dynamical evolution of a galactic disc.

We caution that the assumption of tight-winding (or, WKB limit), which played the crucial role in deriving the dispersion relations in equation~({\ref{stargas-disp}}) and equation~({\ref{onefluid-disp}}), is suspect for the long-wave branch ($|x| < 1$). This is important especially for NGC~2997, where the real solution ($|x|$), i.e., where the line $|s|=|s|_{\rm obs}$ cuts the dispersion relation, is less than 2 (see Fig.~{\ref{fig5}}). To verify the validity of the tight-winding limit in a more quantitative manner for all three galaxies considered here, we calculated the quantity $X$, at a radius $R$ equals to 2R$_{\rm d}$, where $X$ is defined as follows:

\begin{equation}
X=\frac{\kappa^2 R}{2\pi G m \Sigma_{\rm tot}}
\end{equation} 
$\Sigma_{\rm tot}$ denotes the sum of the surface densities of the stellar and the gaseous components and $m$ (= 2, here) denotes the number of spiral arms. For the gas surface density tending to zero, it reduces to the usual $X$, defined for the one-component stellar case, as expected \citep[for details see][]{BT87}. The quantity $X$ denotes the cotangent of the pitch angle of waves of the critical wavenumber $k_{crit}$ and $X >> 1$ implies that the spiral arm is  tightly wound, and hence the tight-winding approximation holds good \citep{BT87}. We followed the following prescription for calculating the quantity $X$.\\
For a given gas fraction $\epsilon$ and for an observed gas surface density at a given radius $R$, we can obtain the total surface density at that radius. Further from the observed rotation curve, the epicyclic frequency $\kappa$ at $R=2R_{\rm d}$ is obtained. The total gas surface density for NGC~6946 is taken from \citet{Cros07} and for M~51, it is taken from \citet{Schu07}. A similar distribution of the total gas surface density for NGC~2997, as a function of radius, is not available in the literature, to the best of our knowledge. So for this case we used the following technique to have an estimate of gas surface density. The $HI$ observations for NGC~2997 give a total $HI$ mass of 4.2$\times$ 10$^9$ M$_\odot$, which extends over an area having mean diameter of 31.2 kpc \citep{Kod11}. This, in turn, gives a mean HI surface density of 5.5 M$_\odot$ pc$^{-2}$ for NGC~2997. Now assuming a constant value of 0.5 for the ratio of mass in HI to mass in H$_2$ \citep[for details see][]{YoSc91}, we get the mean surface density of H$_2$. Then using $\epsilon$, we get the total mean surface density which is used in equation (9) to derive the quantity $X$. The results for value of $X$ for the three galaxies are given in Table~{\ref{table1}}.
\begin{table}
\centering
\caption{Values for $X$ for three galaxies, calculated at $R=2R_{\rm d}$}
\begin{tabular}{ccccccc}
\hline
Galaxy & $R$ & $\kappa$ &$\Sigma_{\rm g}$ & $\epsilon$ & $\Sigma_{\rm tot}$ & $X$\\
name & (kpc) & (km s$^{-1}$  & (M$_{\odot}$  && (M$_{\odot}$ \\
&&kpc$^{-1}$) & pc $^{-2}$) && pc $^{-2}$)\\
\hline
NGC~6946 & 6.6 &42.4 & 22.4 & 0.25 & 89.6 & 2.3 \\
NGC~2997 & 8.0 & 32.7 & 8.3 & 0.15 & 55.3 & 2.7\\
M~51 & 8.0 & 37.1 & 10 & 0.15 & 66.7 & 2.9  \\
\hline
\end{tabular}
\label{table1}
\end{table}
From Table~{\ref{table1}}, we see that the values of $X$ is greater than 1, thus the tight-winding approximation is satisfied, though not by a comfortable margin.

\subsection {A general constraint on pattern speeds}
For a collisionless disc, the density wave can exist only in the regions where
\begin{equation}
\Omega-\kappa/2 \le \Omega_{\rm p} \le \Omega+\kappa/2
\end{equation}
\noindent holds, and the equality holds only at the resonance points.\\
Here in this section, we use our technique based on the calculation of $|s|_{\rm cut-off}$ to constrain the range of allowed pattern speed for the $m=2$, grand-design spiral structures, which will give a stable density wave. The prescription is as follows.

First we consider $Q_s$ values in the range from $1.3$ to $2.0$ and $Q_g$ also in the range from $1.4$ to $1.8$. The gas-fraction, $\epsilon$, is varied from 5 to 25 per cent of the total disc mass, depending on the Hubble-type of a particular galaxy. Then for a fixed value of $\epsilon$, we obtained the value of $|s|_{\rm cut-off}$ for the whole range of $Q_s$ and $Q_g$ considered here. We found that for the gas-fraction of 5 per cent, the gas does not contribute much towards getting a stable density wave, but as the gas-fraction increases steadily, the effect of gas becomes more prominent. However, galaxies having $\epsilon$ $>$ 25 per cent mainly show a flocculent spiral structure and do not generally host a grand-design spiral structure \citep{Elm11}. Considering the above points, we have restricted the value of $\epsilon$ from 10 to 25 per cent.

Now suppose, at a certain radius $R$ for a particular galaxy, one knows the value of $\epsilon$, and one can make a reasonable choice for the values of $Q_{\rm s}$ and $Q_{\rm g}$. Then from these parameters, we can  obtain the value of $|s|_{\rm cut-off}$, call it $\alpha$. 
Then, to get a stable wave, the pattern speed ($\Omega_{\rm p}$) has to satisfy equation (8) with $|s|_{\rm obs}$ being replaced by $|s|$.
This in turn gives
\begin{equation}
s \ge \alpha \hspace{0.5 cm}\mbox{or} \hspace{0.5 cm}s\le -\alpha\,
\end{equation}
\noindent i.e., 
\begin{equation}
\Omega_{\rm p} \ge \Omega+\frac{\alpha \kappa}{2} \hspace{0.3 cm} \mbox{or} \hspace{0.3 cm} \Omega_{\rm p} \le \Omega-\frac{\alpha \kappa}{2}
\end{equation}
depending upon whether one is outside the CR or inside the CR.
Now combining inequality (10) and (12), we can say that when one is inside the CR then the allowed range of pattern speed would be ($\Omega- \kappa/2$, $\Omega-\alpha \kappa/2$) and when one is outside the CR then the allowed range would be ($\Omega+\alpha \kappa/2$, $\Omega+\kappa/2$).

We applied this technique to the three galaxies considered here, at $R=2R_d$, using $Q_{\rm s}=1.5$ and $Q_{\rm g}=1.5$ that were used as a set of parameters in Fig.~{\ref{fig4}}, Fig.~{\ref{fig5}} and Fig.~{\ref{fig6}}. For NGC~6946 and NGC~2997, $R=2R_{\rm d} $ lies inside the CR, and for M~51, $R=2R_{\rm d} $ lies outside the CR. Hence following the above prescription, the predicted range of allowed pattern speeds that give a stable density wave, would be ($\Omega- \kappa/2$, $\Omega-\alpha \kappa/2$) for NGC~6946 and NGC~2997, and for M~51, predicted range of allowed pattern speeds would be ($\Omega + \alpha\kappa/2$, $\Omega + \kappa/2$). The results are summarized in Table~{\ref{table2}}. We check that the observed values of the pattern speed, 22 km s$^{-1}$ kpc$^{-1}$ for NGC 6946, 16 km s$^{-1}$ kpc$^{-1}$ for NGC 2997, and 38 km s$^{-1}$ kpc$^{-1}$ for M~51, do indeed lie within the respective range of allowed pattern speeds obtained theoretically.
\begin{table*}
\centering
\caption{Range of allowed pattern speeds for NGC~6946, NGC~2997 \& M~51 at $R=2R_{\rm d}$}
\begin{tabular}{cccccccc}
\hline
Galaxy & $\Omega$ & $\kappa$ & $\epsilon$ & observed $\Omega_{\rm p}$ & $\alpha$ & Lower bound &  Upper bound   \\
name & (km s$^{-1}$  & (km s$^{-1}$  & (gas  & (km s$^{-1}$&& on $\Omega_{\rm p}$ (km s$^{-1}$  & on $\Omega_{\rm p}$ (km s$^{-1}$   \\
& kpc$^{-1}$) & kpc$^{-1}$) & fraction) &  kpc$^{-1}$)&& kpc$^{-1}$) & kpc$^{-1}$) \\
\hline
NGC~6946 & 30.0 & 42.4 & 0.25 & 22 & 0.17 & 8.8 & 26.4 \\
NGC~2997 & 23.1 & 32.7 & 0.15 & 16 & 0.42 & 6.8 & 16.2 \\
M~51 & 26.2 & 37.1 & 0.15 &  38 &0.42 & 34.0 & 44.8 \\
\hline
\end{tabular}
\label{table2}
\end{table*}

We then varied $Q_{\rm s}$ from 1.3 to 2.0 and $Q_{\rm g}$ from 1.4 to 1.8 to see what ranges of these two parameter will allow the observed pattern speed of three galaxies, to fall in the range, predicted from our theoretical work, and thus give a stable density wave solution. These are summarized in Table~{\ref{table3}}.
\begin{table}
\centering
\caption{Ranges of $Q_{\rm s}$ and $Q_{\rm g}$ for three galaxies, that give a stable wave solution}
\begin{tabular}{ccccc}
\hline
Galaxy & $\epsilon$ & $R (=2R_{\rm d})$ & Range in & Range in \\
name & &(kpc) & $Q_{\rm s}$ & $Q_{\rm g}$ \\

\hline
NGC~6946 & 0.25 & 6.6 &1.3 - 1.7 & 1.4 - 1.8 \\
NGC~2997 & 0.15 & 8.0  &1.3 - 1.5 & 1.4 - 1.8 \\
M~51 & 0.15 & 8.0 &1.3 - 1.9 & 1.4 - 1.8 \\
\hline
\end{tabular}
\label{table3}
\end{table}

We note that a lower value of $Q_{\rm s}$ or $Q_{\rm g}$ than their respective minima chosen here will give an even smaller forbidden region, hence the observed pattern speed would be also permitted for $Q_{\rm s} < 1.3$ and $Q_{\rm g} < 1.4$.

It is interesting to note that for NGC~6946, the allowed range for the pattern speed for a stable
wave is obtained to be 
8.8-17.1 km s$^{-1}$ kpc$^{-1}$ for stars-alone case, while the observed pattern speed 22 km s$^{-1}$ kpc$^{-1}$ clearly lies outside this allowed range.
Similarly, for NGC 2997, the allowed range for the stars-alone case is calculated to be 6.8-13.5 km $^{-1}$ kpc$^{-1}$, while again the observed pattern speed
16 km s$^{-1}$ kpc$^{-1}$ lies outside this permitted range. Thus  real galaxies seem to have pattern speed values that lie beyond the values required for
a stable solution for the stars-alone case. The presence of gas pushes the galaxy to adopt a pattern speed which is higher than the stars-alone case. Thus
the inclusion of gas will have effect in setting the pattern speed and thus in turn will effect the secular evolution of the disc  galaxy (see \S~3.3).

We also illustrate the effect of gas on pattern speeds in another way. Suppose we were to artificially decrease the gas fraction in NGC 6946 to be 15 per cent, then the allowed range for the pattern speeds is obtained to be 8.8-20.8 km $^{-1}$ kpc$^{-1}$ - this does not cover the observed pattern speed of 22 km s$^{-1}$ kpc$^{-1}$. Thus the actual value of gas mass fraction ($\epsilon$) also matters in setting the value of pattern speed. The real galaxies studied seem to have pattern speeds that are close to the 
upper value in the allowed range, or in other words when  $|s|_{\rm obs}$ is just higher than $|s|_{\rm cut-off}$ (see the discussion in Section 3.3).

\section {Discussion}
Here we mention a few points regarding this current work. First, the shape of the dispersion relation and hence the value of $|s|_{\rm cut-off}$ is largely dependent on the values of the parameters ($Q_{\rm s}$,  $Q_{\rm g}$, $\epsilon$) chosen. Note that, for most of the galaxies, the observed radial profiles for $Q_{\rm s}$ and $Q_{\rm g}$ are not available. Consequently, one has to rely on educative guesses for these two Q-values, as we have done here. If the actual observed values for $Q_{\rm s}$ and $Q_{\rm g}$ are known, the findings of this paper can be put in a more accurate way.

Secondly, the density wave corresponding to the observed pattern speed may not be long lasting. In fact, some of the past $N$- body simulations of spiral galaxies have showed that the spiral arms fade out quickly in time \citep{DOBA14}. However when a stellar disc is represented with sufficiently higher number of particles ($\sim$ order of 10$^8$), the spiral structure in the simulations lasts for a time-scale of about a Hubble time, because in these cases the Poisson noise of the system is minimized \citep{Fuji11,Don13}. But, in general, $N$-body simulations of spiral galaxies show that an individual spiral arm is transient and gets wound up, which is at odds with a classical long-lived quasi-stationary density wave scenario \citep{Sel11,Gra12,Ba13}.
 The spiral structures could be more complex, for example,  there is evidence that some galaxies could show either a long-lived spiral pattern or a short-lived pattern, or a galaxy could exhibit both types of patterns, e.g., see the discussion in \citet{Sie12}.

Thirdly, all the results presented in this work, are based on the assumption of the existence of a quasi-stationary density wave that rotates in the galactic disc with a constant pattern speed. However, observational studies by \citet{Foy11} for a sample of 12 nearby late-type star-forming galaxies and by \citet{Fer12} for NGC~4321 did not find any angular off-set in age as predicted by the classical density wave theory. This shows that the density wave theory does not seen to be applicable for these galaxies.

Finally, note that, the results presented here are based on a local calculation, whereas the density wave extends over a large range of radii in the galactic disc. 
The results indicate that gas may have a significant effect on the spiral arms for the large-scale as well. A global modal analysis for a gravitationally coupled two-component (star plus gas) galactic disc to study the effect of the gas on large-scale spiral arms will be taken up in a future study.

\section{conclusion}
In summary, we have shown that the inclusion of gas is essential to get a stable density wave solution corresponding to the observed pattern speed in a spiral galaxy.
We use a two-component (stars plus gas) dispersion relation for a galactic disc, and assume reasonable gas fraction and Toomre $Q$ values for stars and gas. Then we check whether the theoretical dispersion relation permits a real $|k|$ solution (or a stable wave) for the observed rotation curve and the pattern speed.
 Three galaxies of different Hubble type, NGC 6946, NGC 2997, and M~51, are chosen for which the pattern speed values of the grand-design spiral structure and the rotation curves are known from observations. We show that for both NGC~6946 and NGC~2997, at a radius of two disc scale-lengths, the stars-alone case is not able to produce a stable density wave corresponding to the observed pattern speeds, while for M~51, stars-alone case barely supports a stable density wave. One has to include the observed gas fraction in the study, in order to get a stable density wave for these pattern speeds.
 This is the main result from this work. 
 Since these galaxies are typical representatives of their Hubble type, we expect that this finding will hold true for other grand-design spiral galaxies as well.

Based on the technique used here, we obtain a theoretical range of allowed pattern speeds that give a stable density wave at a certain radius of a galaxy. We apply this technique to the three galaxies considered here, and find that the observed pattern speeds of these three galaxies indeed fall in the respective prescribed range.
 
We show that the inclusion of even 15 to 25 per cent gas by mass fraction has a significant dynamical effect on the dispersion relation and hence on the pattern speed which is likely to be seen in a real, gas-rich galaxy. The resulting allowed range of pattern speeds is higher in presence of gas.
The value of pattern speed affects the angular momentum transport. Thus, the gas is likely to play a crucial role in the secular evolution of a galaxy. The effect of gas on the large-scale spiral structure will be investigated in a future study.

\bigskip
\noindent {\bf Acknowledgements:}
We thank the anonymous referee for the constructive comments which have greatly helped to improve the paper. C.J. would like to thank the DST, Government of India for support via 
J.C. Bose fellowship (SB/S2/JCB-31/2014).

\end{document}